\documentstyle[12pt,eqsecnum,multicol,epsfig,aps,prb,array]{revtex}
\newcommand{\bleq}{\ifpreprintsty
		   \else
		   \end{multicols}\widetext \vspace*{-3.5ex}{\tiny
		   
		\noindent\begin{tabular}[t]{c|}
		   \parbox{0.493\hsize}{~} \\ \hline \end{tabular}}
				      \fi}
\newcommand{\eleq}{\ifpreprintsty
		   \else
		   {\tiny\hspace*{\fill}\begin{tabular}[t]{|c}\hline
		    \parbox{0.49\hsize}{~} \\
		    \end{tabular}}\vspace*{-2.5ex}\begin{multicols}{2}
		    \narrowtext
		    \fi}
\newcommand{\bcols}{\ifpreprintsty\else\begin{multicols}{2} 
	\narrowtext\fi}
\newcommand{\ecols}{\ifpreprintsty\else\end{multicols}\fi}

\draft
\widetext
\begin{document}
\title{Rings and rigidity transitions in network glasses} 
\author{Matthieu Micoulaut$^1$ and James C. Phillips$^2$}
\address{$^1$Laboratoire de Physique Th{\'e}orique des Liquides
Universit{\'e} Pierre et Marie Curie,\\ Boite 121
4, Place Jussieu, 75252 Paris Cedex 05, France\\
\ \\
$^2$Dept. of Physics and Astronomy,
Rutgers University, Piscataway, N. J., 08854-8019}

\date{\today}
\maketitle
\begin{abstract}
\par
 Three elastic phases of covalent networks, (I) floppy, (II)
isostatically 
rigid and (III) stressed-rigid have now been identified in glasses at
specific 
degrees of cross-linking (or chemical composition) both in theory and 
experiments. Here we use size-increasing cluster combinatorics and 
constraint counting algorithms to study analytically possible
consequences of 
self-organization. In the presence of small rings that can be locally
I, II 
or III, we obtain two transitions instead of the previously reported
single 
percolative transition at the mean coordination number $\bar r=2.4$, one from a
floppy to 
an isostatic rigid phase, and a second one from an isostatic to a
stressed 
rigid phase. The width of the intermediate phase $\Delta \bar r$ and the
order of the phase transitions depend on
the nature of 
medium range order (relative ring fractions). 
We compare the results to the Group IV chalcogenides, such as Ge-Se
and 
Si-Se, for which evidence of an intermediate phase has been obtained,
and for
 which estimates of ring fractions can be made from structures of high
T crystalline phases.
\par
{Pacs:} 61.20N-81.20P
\end{abstract}
\newpage
\section{Introduction}
A covalently bonded amorphous network progressively stiffens as its
connectivity or mean coordination number $\bar r$ increases\cite{r1,r2}. The
increase of connectivity can be achieved by adding cross-linking 
elements (such as As or Ge) to a starting
chain network of S or Se. From a mechanical
viewpoint, two-fold coordinated single bond chain networks are floppy because the number of nearest
neighbor (r=2) bonding constraints per atom is less than three, the degrees
of freedom\cite{r3}. On the other hand, in the case of networks consisting only
of tetrahedral units (such as amorphous four-fold coordinated silicon), the network is
intrinsically rigid.
\par
In single bond random networks, these simple observations have been described with
success with a mean-field theory based on Maxwell constraint 
counting\cite{r5}. For r-fold coordinateded atoms, the enumeration of atomic constraints 
$n_c^\alpha=r/2$ and $n_c^\beta=2r-3$,
respectively, due to bond stretching and bond-bending forces has shown
that the number of zero-frequency (floppy) modes per atoms actually
vanishes when the mean coordination number\cite{r7,r8} of the network 
increases to the magic number of 2.4. At this point, the network sits
at a mechanically critical point where the number of constraints per
atom $n_c=n_c^\alpha+n_c^\beta$ equals the number of degrees of freedom per
atom.  On the one hand, the exhaustion of all the degrees of freedom means that the network efficiently fills space.  At the same time, because there are no excess constraints, the network can be thought of as globally isostatic. 
\par
Mean-field theory predicts onset of rigidity for $\bar r>2.4$. Numerous
experiments\cite{r9,r10,r11} have confirmed these simple predictions, 
especially in glass
science where bulk chalcogenide glasses have been used as a benchmark
to check these elegant ideas. Threshold behavior has been detected in 
structural\cite{struct}, vibrational\cite{vib}, thermal\cite{therm}
and electronic\cite{electr} properties when $\bar
r$ approaches $2.4$.
Applications of rigidity theory have also been reported in various
fields such as granular matter, biology 
and computational science\cite{Traverse}.  
However, recently it has been shown from Raman scattering and from phase-dependent measurements of the kinetics of the glass transition that two
transitions\cite{PRL97} at $\bar r_{c1}$ and $\bar r_{c2}$ appear
when the network stiffens.
This suggests that the mean-field constraint counting alone (leading to
the single percolative transition) fails to 
describe completely network changes.  These two transitions define an
intermediate phase in which the connected structure continues to be 
stress-free\cite{Selva,JNCS00} (isostatically rigid). 
\par
In this paper, we present a simple way to go beyond the mean-field description of rigidity and include local stress corrections. This is achieved by performing Maxwell constraint counting
on size increasing structures, starting from the short-range level
corresponding to the previous mean-field approach. One main advantage
is that medium range order effects such as small
rings can be taken into account in this construction. 
It appears from this analysis that these small rings mostly determine the nature of the intermediate phase and 
the values of the critical coordination numbers $\bar r_{c1}$ and
$\bar r_{c2}$, hence the width of the intermediate phase $\Delta \bar r=\bar r_{c2}-\bar r_{c1}$. To apply this construction, we choose the simplest case that can be built
up, and which has received considerable attention in the context of
rigidity, namely single-bonded Group IV chalcogenide glasses of the form
$B_xA_{1-x}$ with coordination numbers $r_A=2$ and $r_B=4$ defining the mean coordination number $\bar r=2+2x$. We have used size-increasing cluster approximations (SICA) to construct these size-increasing structures
and  medium range order (MRO) on which
we have realized constraint counting. As hoped, the analysis reveals two transitions: a first one, at which the number of floppy modes vanishes, is closely related to what has been previously obtained in the mean-field approach, and a second one (a ``stress transition''), beyond which stress in the entire structure can no longer be avoided. The second transition can be obtained only under certain conditions which we detail below. We show that the orders of these phase transitions are different and depend on the fraction of ring structures. In between the two transitions, one can define an almost stress-free network structure for which the fraction of isostatic clusters can be computed. 
\par
The paper is organized as follows. In Section II, we show how to construct from small molecules size increasing clusters using SICA and perform Maxwell constraint counting on them. Section III is devoted to the results obtained from the construction, and the change in structure and energy with increasing connectivity. We discuss the results obtained and compare them with
chalcogenide glasses in Section IV and finally, we extend the approach
to fast ionic conducting glasses in Section V.
\section{Size increasing cluster approximations}
\subsection{Construction}
In this section, we describe size increasing cluster approximations
(SICA). This approach has been first introduced to describe the ring
statistics and the intermediate range order in amorphous
semi-conductors\cite{Dina} such as $B_2O_3$ but other applications
have been considered such as the high temperature formation of
fullerenes\cite{C60} and the cell distribution in quasi-crystals\cite{quasi}. In principle, any structural quantity that is computed when one increases the size of a given structure (or a starting network) converges to its ``true'' value if the size becomes almost infinite. In practice, one hopes that the convergence is rapid enough to give reasonable values for medium-sized clusters, yielding information about MRO structures. In quasi-two-dimensional $v-B_2O_3$ clusters having ten boron atoms allows one to obtain a fraction of boron atoms trapped in boroxol rings which is in very fair agreement with experiment\cite{Dina}, but larger clusters may be needed for three-dimensional networks. 
\par
The basic level of the SICA construction is the restricted mean-field approximation where the probability of the short range order structure is derived from the macroscopic concentration, assuming that the cations and anions alternate in the network (chemical ordering). 
 This basic level is
denoted by $l=1$. Then, we construct $l=2$, $l=3$ etc. and compute the
corresponding probabilities in the Canonical Ensemble with
particular energy levels $E_i$ corresponding to bond creation
between the ($l=1$) short range order molecules which are used as
building blocks from step ($l=1$) (corresponding in $Ge_xSe_{1-x}$
also to the reported
mean-field approach\cite{r3}) to arbitrary l. The construction is supposed
to be realized at the
formation of the network, when T equals the fictive temperature $T_f$
so that Boltzmann factors of the form $e_i=exp[-E_i/T_f]$ will be
involved in the probabilities\cite{Galeener}. Since we expect to
relate the width of the
intermediate phase to the ring fraction, we will restrict our present
study to Group IV chalcogenides of the form $Si_xSe_{1-x}$. For the
latter, there is strong evidence that at the stoichiometric
concentration x=0.33 a substantial amount of edge-sharing $SiSe_{4/2}$
tetrahedra\cite{Micol95,r18b}  can be found.

In order to study Group IV chalcogenides, we select basic units such
as the $A_2$ (i.e. $Se_2$) chain
fragment and the stoichiometric $BA_{4/2}$ molecule
(e.g. $GeSe_{4/2}$ which is the majority local structure at
$x=0.333$). These basic units have
respective probabilities $1-p$ and $p=2x/(1-x)$, $x$ being the
macroscopic concentration of the Group IV atoms.
The energy levels are defined as follows. We associate the creation of
a chain-like $A_2-A_2$ structure  with an energy gain of $E_1$, 
$A_2-BA_2$ bondings with an energy gain of $E_2$ and corner-sharing
(CS) and edge-sharing (ES) $BA_{4/2}$ tetrahedra or any ring structure
respectively with $E_3$ and $E_4$. The energy $E_4$ will be used to
change the fraction of edge-sharing relative to corner-sharing tetrahedra. 
The probabilities of the different clusters have statistical weights
$g(E_i)$ which can be regarded as the degeneracy of the corresponding
energy gain and correspond to the number of equivalent ways a cluster
can be constructed. For the step $l=2$, four different clusters can be
obtained (see fig. 1) and their probabilities are given as follows:
\begin{eqnarray}
\label{p1}
p_1&=&{\frac {4(1-p)^2e_1}{4(1-p)^2e_1+16p(1-p)e_2+p^2(16e_3+72e_4)}}
\end{eqnarray}
\begin{eqnarray}
\label{p2}
p_2&=&{\frac {16p(1-p)e_2}{4(1-p)^2e_1+16p(1-p)e_2+p^2(16e_3+72e_4)}}
\end{eqnarray}
\begin{eqnarray}
\label{p3}
p_3&=&{\frac {16p^2e_3}{4(1-p)^2e_1+16p(1-p)e_2+p^2(16e_3+72e_4)}}
\end{eqnarray}
\begin{eqnarray}
\label{p4}
p_4&=&{\frac {72p^2e_4}{4(1-p)^2e_1+16p(1-p)e_2+p^2(16e_3+72e_4)}}
\end{eqnarray}
\par
out of which the concentration of B atoms $x^{(2)}$ can be computed:
\begin{eqnarray}
\label{conc}
x^{(2)}&=&{\frac {p_2+2(p_3+p_4)}{4-p_2-2(p_3-p_4)}}
\end{eqnarray}
Due to the initial choice of the basic units, the energy $E_2$ will
mostly determine the probability of isostatic clusters since the
related Boltzmann factor $e_2$ is involved in the probability
(\ref{p2}) of creating the isostatic $BA_4$ cluster (a $A_2-BA_{4/2}$
bonding). This means that if we choose to
have $E_2\ll E_1,E_3,E_4$, the network will be mainly isostatic in the
range of interest. 
\par 
For the next steps, care has to be taken in order to count the
possible isomers produced from different pathways (e.g. in fig. 1, the
six-membered ring with two B atoms can be produced out of $p_2$ and
$p_3$). More generally, increasing steps will lead to clusters with 
stoichiometry $Ge_nSe_{2l}$ and probability proportional
to $p^n(1-p)^{l-n}$ with $n=0..l$. The corresponding statistical
weights depend much more on the way the clusters are created and have
therefore no general formula depending on $n$ or $l$. However, for the
pure edge-sharing tetrahedra chain, it can be easily checked that its
probability is proportional to $72\times24^{l-2}p^le_4^{l-1}$. Another simple
example is provided by the $Se$ chain whose probability is $4^{l-1}(1-p)^le_1^{l-1}$.
\par
It is obvious that all the cluster probabilites will depend only on two
parameters (i.e. the factors $e_1/e_2$ and $e_3/e_2$) and eventually
$e_4/e_2$ if one considers the possibility of edge-sharing (ES) tetrahedra
or rings. One of the two factors can be made composition dependent since
a conservation law for the concentration of $B$ atoms $x^{(l)}$ can
be written at any step $l$ of the SICA construction\cite{Bray}:
\begin{equation}
\label{1}
x^{(l)}=x
\end{equation}
This means that either the fictive temperature $T_f$ or the energies
$E_i$ depends\cite{Galeener} on $x$ but here only the $e_i(x)$ (or
$e_i(\bar r)$)
dependence is relevant for our purpose. 
\par
The construction has been realized up to the step $l=4$ which already
creates clusters of MRO size.
\subsection{Maxwell cluster constraint counting}
On each cluster one can count Maxwell constraints by enumeration
of bond-bending and bond-stretching constraints and calculation of the
corresponding expressions of $n_c^\alpha$ and  $n_c^\beta$. Of particular
importance are the structures containing a ring (see Fig. 1),
because special care has to be taken to avoid the counting of
redundant constraints\cite{r7}. To illustrate this, let us
consider an isolated
triangle (i.e. a three-membered ring) having a  2-fold atom at each of
its vertices. Of course, this triangle can be completely defined by
three independent variables (e.g. two lengths and one
angle). Performing constraint counting on the atoms will give three 
bond-stretching and three bond-bending constraints, yielding three
redundant constraints. This means that for a three-membered ring, one
has to remove three constraints from the global counting. For the
four-membered ring, this correction is of two constraints, and for a
five-membered ring, of one.
\par
For each step $l$, we have computed the total number of
constraints $n_c^l$:
\begin{eqnarray}
\label{2}
n_c^{(l)}={\frac {\sum_{i=1}^{{\cal N}_l}n_{c(i)}p_i}{\sum_{i=1}^{{\cal N}_l}N_ip_i}}
\end{eqnarray}
where $n_{c(i)}$ and $N_i$ are respectively the number of
constraints and the number of atoms of the cluster of size $l$ with 
probability $p_i$. ${{\cal N}_l}$ is the total number of clusters of
size $l$.
At step $l=2$, it is easy to check that:
\begin{eqnarray}
\label{2b}
n_c^{(2)}&=&{\frac {4p_1+15p_2+22p_3+20p_4}{2p_1+5p_2+6(p_3+p_4)}}
\end{eqnarray}
We have determined either
$e_1/e_2$ or $e_3/e_2$ by solving equ. (\ref{1}), and once
these factors become composition dependent, it is possible to compute
the probabilities $p_i$ as a function of composition and find
for which concentration $x$ (or which mean coordination number $\bar
r$) the system reaches optimal glass formation where the number of
floppy modes $f_l=3-n_c^{(l)}$ vanishes.
\section{Results}
\subsection{Structural properties}
In this section, we consider the solutions of the SICA construction
under various structural possibilities. 
\par
The simplest case is
random bonding, which is obtained when the cluster
probabilities $p_i$ are only given by their statistical weights
$g(E_i)$. This would for example reduce the probability $p_4$ in
equ. (\ref{p4}) to  
\begin{eqnarray}
\label{p4r}
p_4&=&{\frac {18p^2}{(1+p)^2+18p^2}}
\end{eqnarray}
with $p=2(\bar r-2)/(4-\bar r)$.
A single solution is obtained for the glass optimum point defined by
the vanishing of the number of floppy modes
$f=0$ at all SICA steps, in the mean coordination number range 
[2.231,2.275], slightly lower than the usual mean-field value of 2.4.
Since there is only one solution, there is no intermediate  phase in the case of random
bonding.
\par
Self-organization of the network can be obtained by starting from a floppy
cluster of size $l$ (e.g. a chain-like structure made of a majority of
A atoms), and allowing the agglomeration of a new
basic unit onto this cluster to generate the cluster of size $l+1$ 
only if the creation of a stressed rigid region can be avoided
on this new cluster. This happens when  two $BA_{4/2}$ basic
units are joined together on a given cluster.
With this rather simple rule, upon increasing $\bar r$ one
accumulates isostatic rigid regions on the size increasing clusters
because $BA_{4/2}$ units are only accepted in 
$A_2-BA_{4/2}$ isostatic bondings with energy $E_2$. On the opposite side, one
can start at high concentration, close to the mean coordination number
of $\bar r=2.67$  and
follow the same procedure but in opposite way, i.e. with adding A
atoms, one allows only
bondings which lead to isostatic rigid or stressed rigid regions, excluding
systematically the possibility of floppy $A_2-A_2$ bondings.
\par
In the case of self-organized clusters, the simplest case to be
studied is the case of dendritic clusters, where no rings are allowed
(achieved by setting $e_4$ to zero). For an infinite size $l$, this
would recover the results from Bethe lattice solutions or  
Random Bond Models\cite{RBM} for which rings are also excluded in the
thermodynamic limit\cite{Bethe}. A single transition for even $l$
steps at exactly the
mean-field value $\bar r=2.4$ is obtained whereas for the step $l=3$, there is
a sharp intermediate phase defined by $f=0$ (still at $\bar r=2.4$) and the 
vanishing of floppy regions (i.e. $e_1/e_2$ is zero) at $\bar
r=2.382(6)$. Once the probabilities of floppy, isostatic rigid and stressed
rigid clusters as a function of the mean
coordination number are computed, it appears 
that the network is entirely isostatic at the point where $f=0$
(solid line, fig. 2). There the number of degrees of freedom per atom
is exactly three.
\par
The intermediate phase shows up if a certain amount of medium range
order (MRO) is allowed. This is realized in the SICA construction by
setting the quantity $e_4/e_2$ non zero, i.e. edge-sharing (ES) 
tetrahedra $BA_{4/2}$
leading to four-membered rings $B_2A_4$ can now be created at the growing
cluster steps. This means also that if stress should be created when
$\bar r$ is increasing, then it should be only in ring structures and
not by two corner-sharing connected $BA_{4/2}$ tetrahedra.
\par
Two transitions are now obtained for
every SICA step. The first one lies always around the mean coordination
number $\bar r_{c1}=2.4$ where the number of floppy modes $f$
vanishes. The second transition is located
at $\bar r_{c2}$ and is a new feature of rigidity theory. When
starting from a floppy network and progressively stiffening the
network and requiring self-organization, the network will reach a
point beyond which stressed 
rigid bondings outside of ring structures can not be avoided anymore. 
This point is a stress transition. We show in fig. 2
the $l=2$ result where $f=0$ at $\bar r_{c1}=2.4$ for different 
fractions of ES tetrahedra, defining an
intermediate phase $\Delta \bar r$. We should also stress that even for a
non-zero ES fraction, $f=0$ is always obtained at $\bar r=2.4$ for
$l=2$. From this analysis, it appears that 
the first transition at $\bar r_{c1}$ does not depend on the ES
fraction, as well as the fraction of stressed rigid clusters
in the structure. In fig. 2, the probabilities of the related stressed rigid
clusters for a non-zero ES fraction can of course be obtained from the
floppy and isostatic ones since the sum of all probabilities is equal to one.
\par
To ensure continuous deformation of the network when
B atoms are added while keeping the sum of the probability of floppy, isostatic
rigid and stressed rigid clusters equal to one, the probability of isostatic 
rigid clusters connects the isostatic solid line at $\bar r_{c2}$. 
Stressed rigid rings first appear in the region $\bar r_{c1}<\bar r<\bar r_{c2}$
while chain-like stressed clusters (whose probability is proportional
to $e_3$) occur only beyond the stress
transition, when $e_3\neq0$. This means that within this approach, when 
$\bar r$ is increased, stressed
rigidity nucleates through the network starting from rings, as ES
tetrahedra or small rings. It is easy to see from fig. 2 that
the width $\Delta\bar r=\bar r_{c2}-\bar r_{c1}$ of
the intermediate phase increases with the fraction of ES. This can be
extended to any MRO fraction (fig. 3) and it shows 
that $\Delta \bar r$ is almost an increasing function of the ES
fraction as seen from the result at SICA step $l=4$.  Since there is only a
small difference between allowing only four-membered rings (ES) (lower
dotted line) or
rings of all sizes (upper dotted line) in the clusters, we conclude
that the ES mostly determine the stress transition and hence the
width. 
\par
Finally, one can see from
fig. 2 and the insert of fig. 3 that the probability of isostatic clusters is
maximum in the window $\Delta \bar r$, and almost equal to 1 for
the even SICA steps, providing evidence that the molecular structure
of the network in the window is almost 
stress-free. The point at which $f=0$ shifts slightly around $\bar
r=2.4$ with the SICA step $l$ (e.g. see the insert of fig. 3).
\subsection{Constraint free energy}
From the cluster distribution obtained by the SICA, it is
also possible to compute the constraint-related free energy,
following the approach reported by Naumis\cite{Naumis}. Here, we have
kept from the internal energy of the network only the part related to
energy of the elastic deformations of the network, removing the
contributions from the harmonic vibrations of the atoms and the
anharmonic contributions which are irrelevant for our purpose 
(however, see \cite{Naumis}). This idea is also consistent with the
work of Duxbury and co-workers who showed that the number of floppy
modes behaves as a free energy for both rigidity and connectivity
percolation\cite{r22b}. The entropy of the network can be
evaluated as a Bragg-Williams term from the distribution of cluster 
probabilities $p_i$ at step $l$.
\begin{eqnarray}
\label{entro}
F_l=U_l-TS_l=Nk_BTf_l+Nk_BT\sum_{i=1}^{{\cal N}_l}p_i\ln{p_i}
\end{eqnarray}
where $f_l=3-n_c^{(l)}$ is the number of floppy modes computed following
equ. (\ref{2}). Since we have expressed the latter quantity as a
function of the mean coordination number $\bar r$ and since the
probabilities can also be expressed as a function of $\bar r$, the
free energy can be plotted as a function of $\bar r$ in the
case of self-organization. Figure 4 shows the constraint related free
energy $F_l$ for two non-zero ES fractions. 
\par
It appears from the figure that the second transition at $\bar =\bar r_{c2}$
(the ``stress transition'') is a first order transition while the
first transition at $\bar r_{c1}$ is weakly second order. Moreover, as
one can see from the insert, the first transition at $\bar r_{c1}$ progressively
becomes first order in character when the rate of edge-sharing
tetrahedra $\eta$ is increased. On the other hand, the jump of
$F_l^{(1)}$ at $\bar r=\bar r_{c2}$ is reduced when the ES fraction is 
increased (insert of fig. 4).
\par
Both curves show a marked minimum of $F_l$ in the range [$\bar
r_{c2}$,$2.667$] at a certain coordination number $\bar r_e$ which
signals an equilibrium state with respect to
cross-linking. A major consequence of this result is that one may
expect phase separation in the stressed rigid region leading for
increasing cluster sizes to nano-scale phase separation in the network
backbone. Close to $\bar r=2.667$, the structure should therefore be
made of B-poor clusters having the statistics of the local $F_l$
minimum but also compensating B-rich clusters in order to still
satisfy equ. (\ref{1}). This structural change is driven by the
entropic term appearing in equ. (\ref{entro}) since the energy of the
elastic deformation of the network is zero in the stressed rigid
phase. With increasing ES fraction, this equilibrium state
shifts to the value 2.667. For a ES fraction of 1, equilibrium state
and stoichiometric composition merge together.
Experimental evidence of nanoscale phase separation is
discussed in the following.
\section{Discussion}
\subsection{The Boolchand Intermediate phase}
As mentioned above, chalcogenide glasses are the first systems that
have been carefully studied  and the intermediate phase defined by the
two transitions has been discovered by Boolchand in
the context of self-organization\cite{PRL97,Selva}. 
SICA provides therefore a benchmark to check the results obtained. 
To be specific, Raman scattering \cite{PRL97,Bool1} probes elastic 
thresholds in binary $Si_xSe_{1-x}$ or $Ge_xSe_{1-x}$. The germanium or
silicon corner sharing mode chain frequencies change with mean coordination number $\bar r$ of the glass network. These
frequencies exhibit not only a change in slope at
the mean coordination number $\bar r_{c1}=2.4$, but also a first order jump
at the second transition $\bar
r_{c2}$. In germanium systems, the second transition is located around
the mean coordination number of $2.52$ whereas $\bar r_{c2}=2.54$ in Si
based systems. For both systems, a power-law behavior in $\bar
r-\bar r_{c2}$ is detected (see fig. 5) for $\bar r>\bar r_{c2}$ and the
corresponding measured exponent is very close to the one obtained in
numerical simulations of stressed rigid networks\cite{Tersoff}.
Moreover, these results clearly correlate with the vanishing between $\bar r_{c1}$ and $\bar r_{c2}$ of the
non-reversing heat flow $\Delta H_{nr}$ (the part of the heat flow
which is thermal history sensitive) in MDSC measurements\cite{PRL97,Bool1}.
\par
The study of stoichiometric compounds such as $SiSe_2$ or
$GeSe_2$ also leads to better understanding of medium range
order. In the former, $^{29}Si$ NMR experiments have shown that most
of the tetrahedra were part of long edge-sharing chains\cite{r261} in the
glass. $SiSe_2$ has different crystalline polymorphs which all exhibit
a strong edge-sharing tendency\cite{r262}. The high temperature phase 
is made of $100\%$ edge-sharing tetrahedral\cite{r263}, while different phases 
display a distribution in
terms of NMR $E^{(k)}$ functions (where the subscript $k=0,1,2$ refers
to the number of tetrahedra shared by edges on a tetrahedron), but
with a majority of $E^{(2)}$ structures\cite{r262,r264}. In the
$SiSe_2$ glass, the
fraction of $E^{(2)}$ has been found to be of the order\cite{r265} of $0.53$. On
the other hand,  low temperature crystalline $GeSe_2$ has no
edge-sharing tetrahedral\cite{r266}, but glassy $GeSe_2$ exhibits a 
companion Raman line associated with edge-sharing tetrahedra \cite{r5}).
\par

The SICA approach has shown that the width $\Delta\bar
r$ of the intermediate phase increases mostly with the fraction of ES
tetrahedra. We stress that the width should converge to a lower limit
value of $\Delta \bar r$ compared to the step $l=2$, therefore one can
observe the shift downwards when increasing $l$ from 2 to 4. This
limiting value is in principle attained for $l\to \infty$, or at least for much
larger steps\cite{Micoul} than $l=4$. For Si-Se, $\Delta\bar r=0.14$ is
much more sharply defined than for Ge-Se ($\Delta\bar r=0.12$) consistent with the
fact that the number of
ES tetrahedra is higher in the former\cite{Selva}. 
\subsection{Nanoscale phase separation}
Phase separation effects in glasses exhibit usually pronounced changes
in physical properties and most studies have focused on thermally
driven heterogeneities which can in some cases display bimodal glass
transition temperatures. Here the separation effect results from a change in
network connectivity which has its origin in the free energy minimum
at $\bar r_e$.
\par
It appears that these nano-scale phase separation have been revealed
from compositional trends\cite{Bool1} of the glass transition
temperature $T_g$ because they display a maximum in $T_g$ close to the
stoichiometric concentration. Such a feature has been observed in
Ge-Se\cite{PRL97}, As-Se\cite{KasapAsSe} alloys, but not in
Si-Se\cite{Selva}. These $T_g$ trends can be compared with
spectroscopic (Raman\cite{Raman}, M{\"o}ssbauer\cite{Bool1}) data which also
give evidence of broken chemical order, suggesting that the structure
of stoichiometric glasses such as $GeSe_2$ or $As_2Se_3$ is made of
a chalcogen rich majority phase and a compensating Ge- or As- rich
phase. Furthermore, in the metal rich phase Ge-Ge or As-As bonds are present.
\par
The difference between the Si-Se and the Ge-Se glass lies in the
following. Since the Si-Se has a much higher ES fraction compared to
Ge-Se, the value of its corresponding local constraint free energy
equilibrium will lie very close to the value $\bar r=2.667$ (see
insert of fig. 4). The Ge counterpart will have the same minimum at a
lower value in the range [$\bar r_{c2}$, 2.667] because of fewer ES 
tetrahedra thus favouring the emergence of the chalcogen-rich phase 
when $\bar r$ is increased. Our last comments brings us back to
constraint counting. Since Si-Se is more weakly constrained than Ge-Se due
to the higher amount of ES\cite{r7}, the glass transition temperature
of the stoichiometric glass will be higher compared to Ge-Se.
\section{Application to fast ionic conductors}
One interesting field of application of cluster construction and
constraint counting algorithms is the field of fast ionic conductors
(FIC)\cite{FIC1,FIC2}, which has received
considerable attention in the last fifteen years because of
potential applications of these solid electrolytes in all solid state
electrochemical devices and/or miniaturized systems such as solid state
batteries. 
An important step forward has been made by replacing the oxygen in
usual oxide glasses by more polarizable chalcogenide atoms (S,Se
mostly) which has increased the dc conductivity\cite{condux1}
in these systems by several orders of magnitude\cite{condux2}, up to a
value of about $10^{-3}~\Omega^{-1}.cm^{-1}$.
Surprisingly, the extension of constraint theory from network
chalcogenide glasses such as $As_xSe_{1-x}$ to ionic glasses has
received little attention and to our knowledge, has been only reported
for a few oxide glasses\cite{Science,SSC}. Elastic
percolative effects in these types of networks have not been studied so
far with the network change in solid electrolytes, although it is certainly
fundamental for 
the understanding of the mobile carriers' motion since various models 
of conductivity\cite{r341,r342} stress the importance of the mobility
$\mu$ in the
contribution to conductivity. Obviously, the mobility is related to a
local mechanical deformation of the network\cite{Barrio}, allowing a
cation to move through holes in the structure. In terms of rigidity, 
one may therefore expect that the mobility $\mu$ in a stressed rigid
solid electrolyte should be substantially lower compared
to the cation mobility in a floppy one, because in the latter floppy
modes allow a local  low energy deformation. The percolative effect of
mobility should certainly show up in this kind of network so that the
conductivity $\sigma$ should display some particular behavior in the
stress-free intermediate phase and at the two transitions.

\par
The SICA approach can be applied to the present solid electrolyte case
by considering the simplest binary conducting glass, which is of the
form $(1-x)SiX_2-xM_2X$ with X an anion of Group VI (X=O,S,Se) and M an
alkali cation (M=Li,Na,K,\ldots). The free carriers are the $M^\oplus$
cations. The local structure in these glasses can be determined by
many different expertiments and is usually described  
in terms of so-called $Q^4$ and $Q^3$ units, derived from NMR
data\cite{Qn}. The former corresponds to the usual silica tetrahedron made of
one silicon and four Group VI atoms at the corner [e.g. $SiSe_{4/2}$] 
while the latter has one additional anion bonded to the alkali cation
[e.g. $SiSe_{5/2}^\ominus Na^\oplus$]  that
does not connect anymore to the network
(fig. 6). Although it is yet not clear what
is the coordination number of the alkali cation\cite{r343}, it can be assumed
that the strongest interaction of the alkali cation is the one related
to the NBO. This means that the effective coordination number of $M$ is taken as
one, as suggested by different authors\cite{SSC,r344}.
\par
Starting from the local $Q^4$ and $Q^3$ units (with respective
probabilities $1-p$ and $p=2x/(1-x)$, the SICA probabilities can be 
evaluated for different steps of
cluster sizes following the procedure described previously and taking
into account the 1-fold M cations\cite{Bool2}. When
constraint counting is performed, it appears 
that the creation of a $Q^4-Q^4$ connections leads
to a stressed rigid cluster, while the $Q^4-Q^3$ and $Q^3-Q^3$
connections yield
respectively isostatically stressed and floppy clusters. The SICA
results show again that the intermediate phase exists only when a non-zero
fraction of small rings is allowed in the self-organized structure.
The corresponding results are displayed in 
fig. 7 for $l=2$ and work on higher SICA steps is in
progress\cite{CMV}. Rigidity nucleates here in a way opposite to network
chalcogenides.  The vanishing of the number of floppy modes defines
the upper limit of the intermediate phase, while stressed rigidity
outside of ring structures disappears for $x>x_{c1}$. This is
consistent with the fact that the network is stressed rigid at low modifier
concentration. However, the base network glass is stress free. 
In $SiO_2$, the Si-O-Si angle distribution is quite wide leading to
broken bond-bending constraints on oxygen\cite{r346}, while in the $SiS_2$ and
$SiSe_2$ glasses, the structure is mostly made out of edge-sharing
$SiSe_{4/2}$ or $SiS_{4/2}$ tetrahedra\cite{r261} which are weakly stressed
(i.e. $n_c\leq 3.667$ a value which would be expected in a dendritic
network at the mean coordination number $\bar r=2.667$ or
concentration $x=0.333$) due to
the 4-membered ring correction coming from the counting of redundant
constraints\cite{r7}.
\par
Increase of the alkali content leads to an increase of 
floppiness. In the oxide system $(1-x)SiO_2-xM_2O$, the width
$\Delta x$ should be very small or zero since the fraction of ES tetrahedra
in the oxide systems is almost zero\cite{Galeener}. Still, percolative
effects are expected
at the concentration $x=0.2$ corresponding to the transition from
rigid to floppy networks, a transition that has been observed in
sodium tellurate glasses\cite{Science}. In sulfur and selenide glasses such as 
$(1-x)SiS_2-xNa_2S$, the width should be much broader because of the
existence of the high amount of edge-sharing tetrahedra in the $SiS_2$
or $SiSe_2$ base networks\cite{r265}. For glasses with a high
amount of ES tetrahedra, the lower limit at $x_{c1}$ of the intermediate phase is
expected to
decrease down to $x=0$ for the limiting case $\eta=1$.
In the sulfur base glass, $^{29}Si$ NMR
have shown that the fraction of ES tetrahedra should be about 0.5, 
slightly higher than in the selenide analogous system\cite{ES1,ES2}. 
\par
From fig. 7, for $\eta=0.5$ one
should observe
a window of about $\Delta x=0.09$. Unfortunately, conductivity, structural
and thermal results\cite{Pradel} on
these systems
are only available for an alkali concentration
$x>0.2$. However, in the
different silica based glasses, a rigidity transition has been 
observed\cite{Vaills} at
the concentration $x=0.2$ which should provide guidance for
forthcoming studies in this area.
\par
Finally, temperature effects should
be observable close to this transition. Since the concentration of alkali free
carriers $n_L$ depends on the temperature (the higher the temperature, the
higher $n_L$), an increase of the temperature $T$ should lead to a
decrease of the number of network constraints, the fraction of intact
bond-stretching constraints $n_c^\alpha$ of the alkali atom being
proportional to $1-n_L$. Consequently, a shift of the mechanical
threshold ($f=0$) to the higher concentrations should result from an
increase of T.
\section{Summary and conclusions}
To summarize, we have shown in this article how stress change in  
molecular systems can be described using both cluster construction and
constraint counting. This permits to go beyond the usual mean-field
approach of rigidity and to obtain the two observed rigidity
transitions. We have found that there is a single transition
from floppy to rigid networks in a certain number of structural
possibilities. An intermediate phase appears when a fraction of
medium range order is allowed in self-organized networks and the order
of the underlying phase transitions is first and second
order, and depend also on the fraction of ES. Nano-scale phase
separation appears in the stressed rigid phase and is driven by the
cluster entropy. This separation leads to Group VI-rich clusters and
chalcogen rich clusters when the stoichiometric composition is
attained.
Finally, extension of this approach to ionic conductors has
been emphasized and should motivate new developments in this field.
\par
LPTL is Unit{\'e} Mixte de Recherche associ{\'e}e au CNRS n. 7600


\newpage
\listoffigures
\newpage
\begin{figure}
\begin{center}
\epsfig{figure=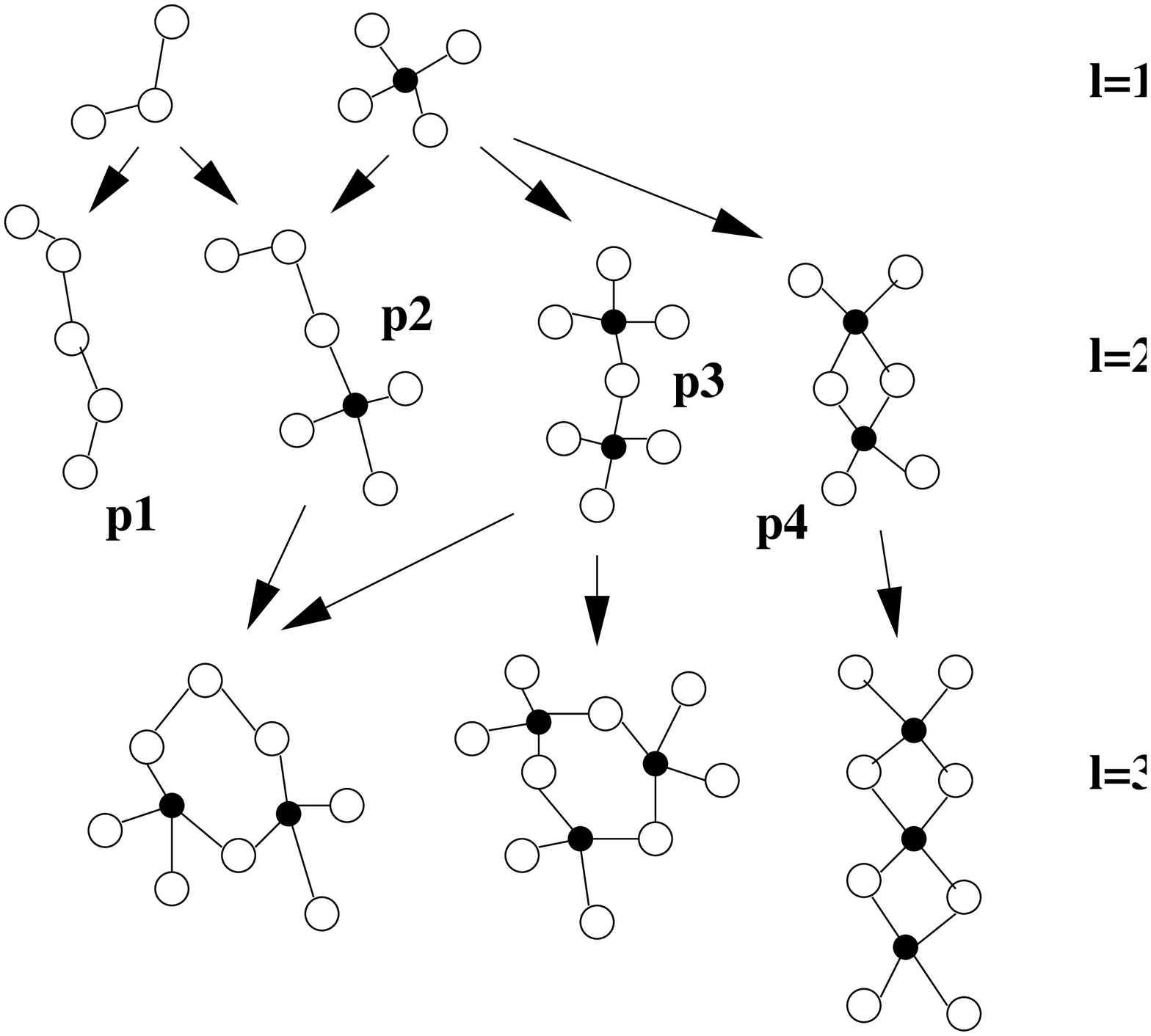,width=12cm}
\caption{From the short range order molecules ($l=1$) yielding the mean-field
result to all clusters at $l=2$ and some MRO produced at $l=3$. Isomers
start to be created at step $l=3$. Note the creation of medium range
order such as edge-sharing tetrahedra or rings. Each boundary atom is
counted half.}
\end{center}
\end{figure}
\newpage
\begin{figure}
\label{rate}
\begin{center}
\epsfig{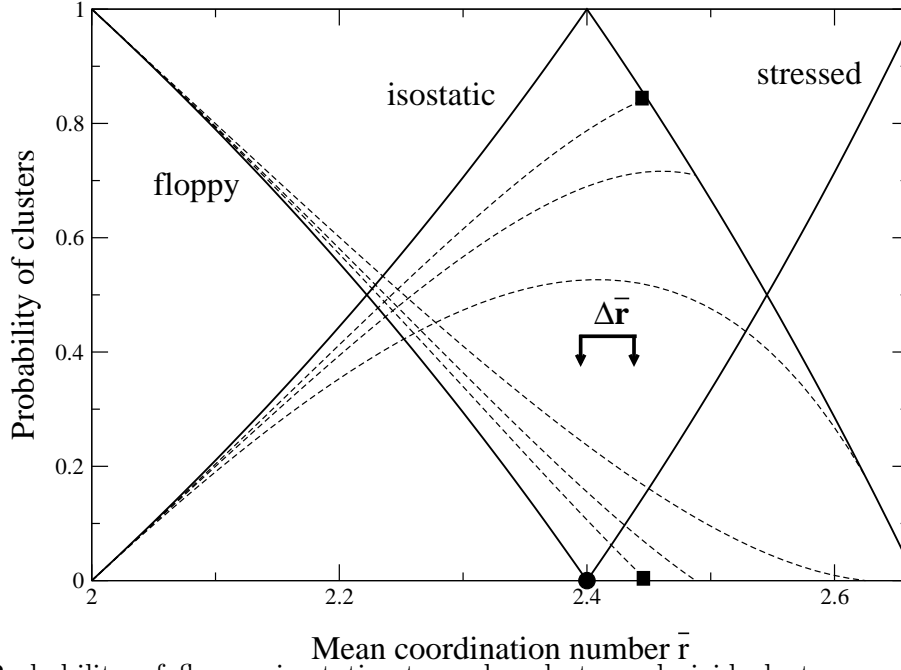}
\caption{Probability of floppy, isostatic stressed and stressed rigid
clusters, as a function of the mean coordination number $\bar r$ for
different possibilities of medium range order. The solid line
corresponds to the dendritic case while the broken lines correspond to
a respective ES fraction at the stress transition of 0.156, 0.290 and
0.818. The filled square indicates the stress transition at the point
$\bar r_{c2}$ and the filled circle the point $\bar r_{c1}$ that does
not depend on the ES rate (see text for details). For clarity, we have
removed the probabilities of stressed rigid clusters for non-zero ES 
fractions.}
\end{center}
\end{figure}
\newpage
\begin{figure}
\label{width}
\begin{center}
\epsfig{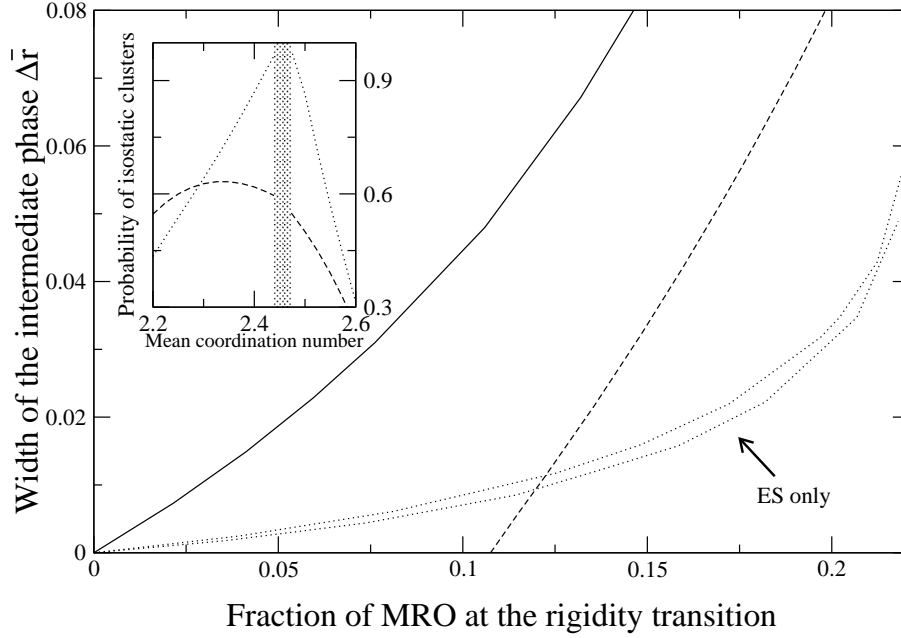}
\caption{Width of the intermediate phase as a function of the fraction
of medium range order at the stress transition for $l=2$ (solid line),
$l=3$ (dashed line) and $l=4$ (dotted lines). The lower dotted line
corresponds to a structure where only ES tetrahedra have been
allowed. The insert shows the probability of isostatic clusters with
mean coordination number for $l=3$ (dashed line) and $l=4$ (dotted
line), compared to the shaded region defined by the $\Delta\bar r$ from
SICA analysis. The point defined by $f=0$ is shifted compared to $l=2$.}
\end{center}
\end{figure}
\newpage
\begin{figure}
\label{free}
\begin{center}
\epsfig{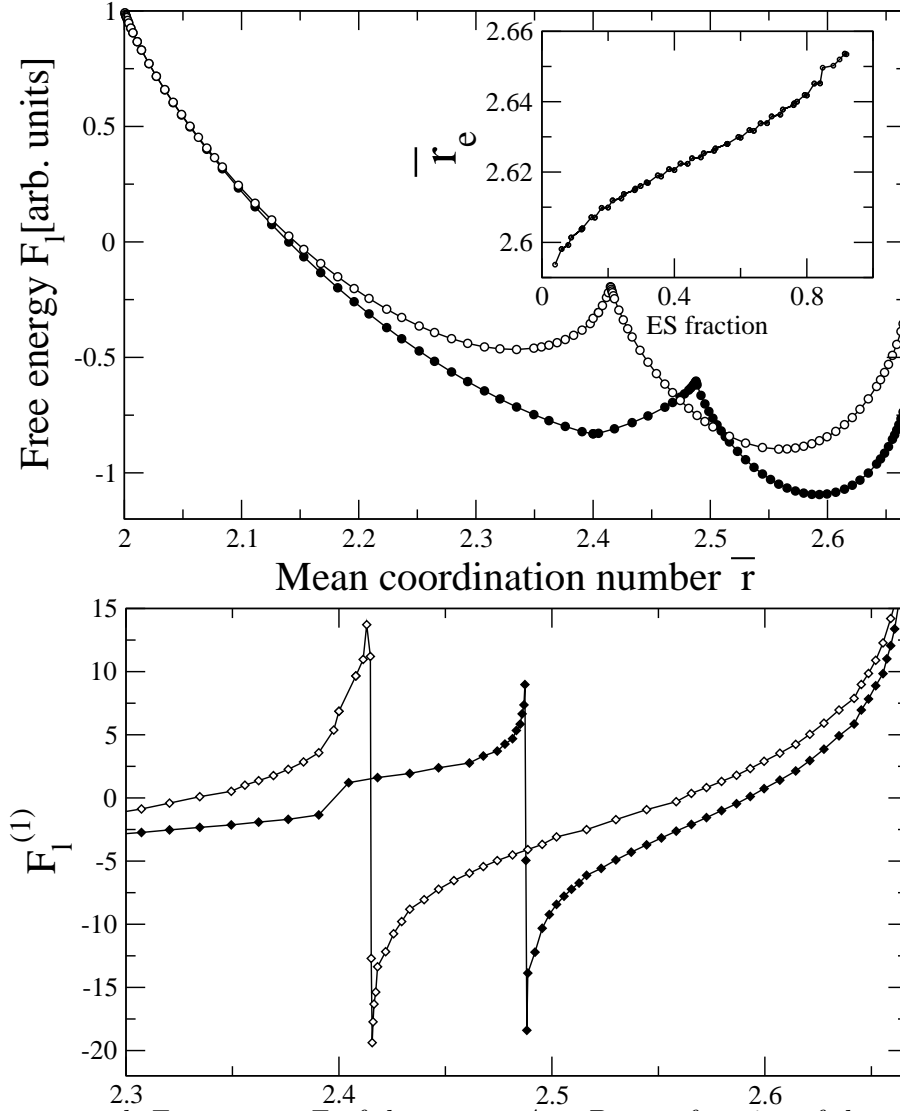}
\caption{Upper panel: Free energy $F_l$ of the system $A_{1-x}B_x$ as a function
of the mean coordination number $\bar r$ for different fractions of
edge-sharing units $\eta$. Open circles: $\eta=0.29$, filled circles:
$\eta=0.56$. The insert shows the equilibrium coordination number $\bar
r_e$ with respect to the ES fraction (see text for details). Lower
panel: the first derivative of the free energy $F_l^{(1)}$ with
respect to $\bar r$ as a function of $\bar r$.}
\end{center}
\end{figure}
\newpage
\begin{figure}
\label{nucs}
\begin{center}
\epsfig{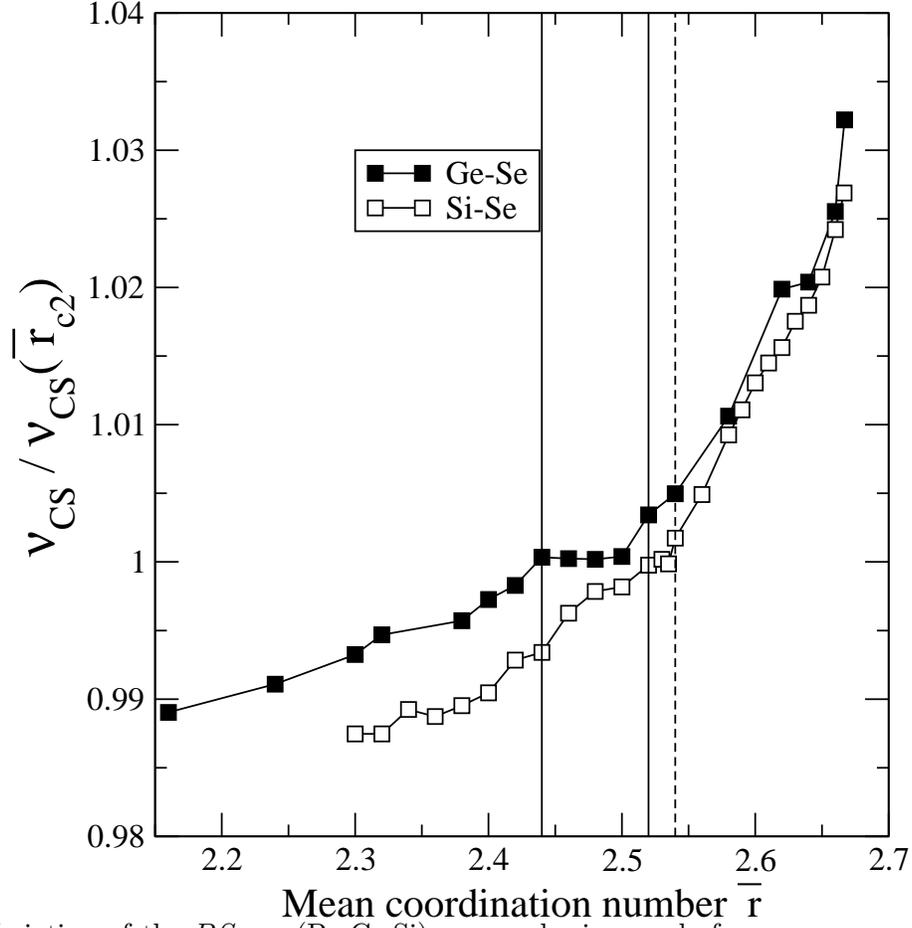}
\caption{Variation of the $BSe_{4/2}$ (B=Ge,Si) corner-sharing mode
frequency normalized to $1$  in Raman spectroscopy with respect to the
mean coordination number $\bar r$. The solid vertical lines define the
intermediate phase in Ge-Se\cite{PRL97} while the lower solid line and
the dashed line define it for Si-Se, following \cite{Selva}. The Si
Intermediate phase is larger than the Ge one.}
\end{center}
\end{figure}
\newpage
\begin{figure}
\label{tetra}
\begin{center}
\epsfig{figure=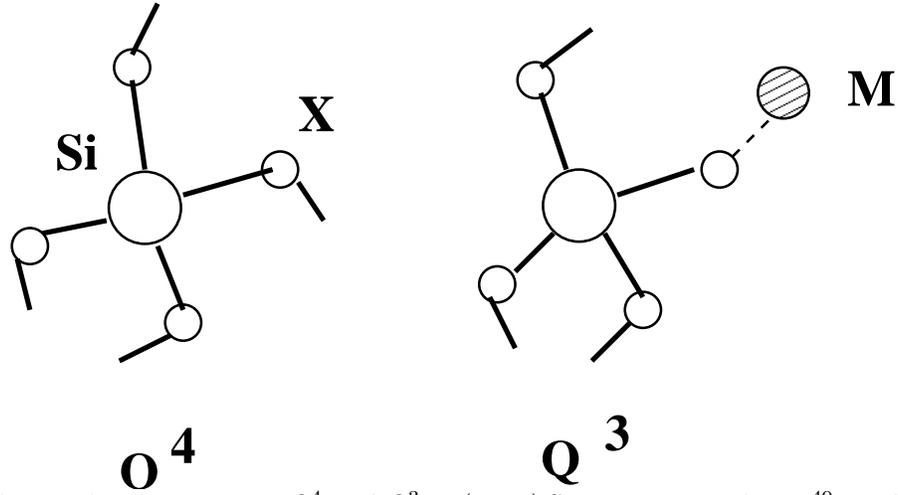,width=12cm}
\caption{The two local structures $Q^4$ and $Q^3$ in
$(1-x)SiX_2-xM_2X$ glasses\cite{Qn}, with $X=O,S,Se$ and
$M=Na,K,\ldots$. The $Q^3$ structure has one non-bridging anion.}
\end{center}
\end{figure}
\newpage
\begin{figure}
\label{widthq}
\begin{center}
\epsfig{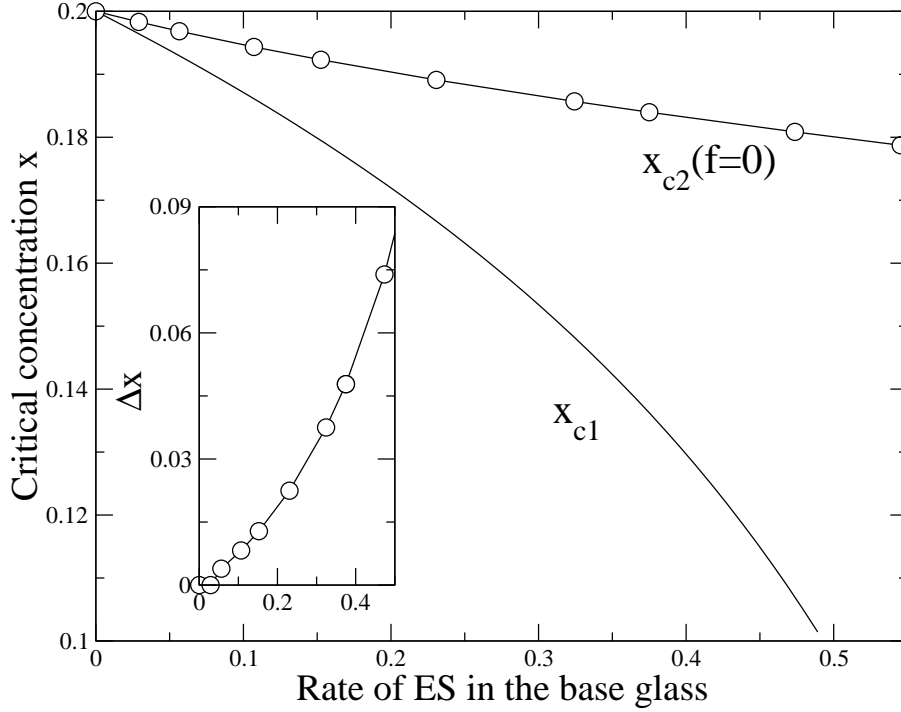}
\caption{Critical concentrations in 
$(1-x)SiX_2-xM_2X$ glasses, with $X=O,S,Se$ and $M=Na,K,\ldots$, as a
function of the fraction of edge-sharing (ES) in the base $SiX_2$
glass. The insert shows the corresponding width of the intermediate
phase as a function of the same quantity.}
\end{center}
\end{figure}

\begin{thebibliography}{00}
\bibitem{r1} W.J. Bresser, P. Suranyi and P. Boolchand,
Phys. Rev. Lett. {\bf 56} (1986) 2493
\bibitem{r2} P. Boolchand, Phys. Rev. Lett. {\bf 57} (1986) 3233
\bibitem{r3} J.C. Phillips, J. Non-Cryst. Solids {\bf 34} (1979) 153
\bibitem{r5} J.C. Phillips, J. Non-Cryst. Solids {\bf 43} (1981) 37
\bibitem{r7} M.F. Thorpe, J. Non-Cryst. Solids {\bf 57} (1983) 355
\bibitem{r8} H. He and M.F. Thorpe, Phys. Rev. Lett. {\bf 54} (1985)
2107
\bibitem{r9} W.A. Kamitarahara, R.L. Cappelletti, P.Boolchand,
B. Halfpap, F. Compf, D.A. Neumann, H. Mutka, Phys.Rev. B {\bf 44}
(1991) 94 
\bibitem{struct} S.S. Yun, H. Li, R.L. Cappelletti, P. Boolchand,
R.N. Enzweiler, Phys. Rev. B{\bf 39} (1989) 8702
\bibitem{vib} K. Murase, T. Fukunaga, in {\em Defects in glasses},
Mater. Res. Soc. Symp. {\bf 61} (1986) 101
\bibitem{therm} U. Senapati, A.K.Varshneya, J. Non-Cryst. Solids {\bf
185} (1995) 289
\bibitem{electr} S. Asoka, E.S.R. Gopal, G. Parathasarathy,
Phys. Rev. B{\bf 35} (1987) 8369
\bibitem{r10} M. Tatsumisago, B.L. Halfap, J.L. Green, S.M. Lindsay
and C.A. Angell, Phys. Rev. Lett. {\bf 64} (1990) 1549
\bibitem{r11} U. Senapati and A.K. Varshneya, J. Non-Cryst. Solids
{\bf 185} (1995) 289; T. Wagner and S. Kasap, Phil. Mag. B{\bf 74}
(1996) 667
\bibitem{Traverse} see {\em Rigidity theory and applications},
M.F. Thorpe, P.M. Duxbury Eds. Fundamental Materials Research Series,
Plenum Press/Kluwer Academic  1999
\bibitem{PRL97} X. Feng, W.J. Bresser and P. Boolchand,
Phys. Rev. Lett. {\bf 78} (1997) 4422
\bibitem{Selva} D. Selvenathan, W.J. Bresser, P. Boolchand,
Phys. Rev. B{\bf 61} (2000) 15061
\bibitem{JNCS00} M.F. Thorpe, D.J. Jacobs, M.V. Chubynsky and
J.C. Phillips, J. Non-Cryst. Solids {\bf 266-269} (2000) 859
\bibitem{Dina} M. Micoulaut, R. Kerner and D.M. dos Santos-Loff,
J. Phys. Cond. Matt. {\bf 7} (1995) 8035
\bibitem{C60} R. Kerner, K. Penson and K.H. Bennemann,
Europhys. Lett. {\bf 19} (1992) 363
\bibitem{quasi} R. Kerner and D.M. dos Santos-Loff, Phys. Rev. B{\bf
37} (1988) 3881
\bibitem{Galeener} F.L. Galeener, D.B. Kerwin, A.J. Miller and
J.C. Mikkelsen Jr., Phys. Rev. B{\bf 47} (1993) 7760
\bibitem{Micol95} M. Micoulaut, Physica B{\bf 212} (1995) 43
\bibitem{r18b} Z. Zhang and J.H. Kennedy, Solid St. Ionics {\bf 38}
(1990) 217
\bibitem{Bray} P.J. Bray, S.A. Feller, G.E. Jellison and Y.H. Yun,
J. Non-Cryst. Solids {\bf 38-39} (1980) 93
\bibitem{RBM} D.J. Jacobs and M.F. Thorpe, Phys. Rev. Lett. {\bf 80}
(1998) 5451
\bibitem{Bethe} M.F. Thorpe, D.J. Jacobs and M.V. Chubynsky in {\em
Rigidity theory and applications}, edited by Plenum press/Kluwer
Academic 1999
\bibitem{Naumis} G.G. Naumis, Phys. Rev. B{\bf 61} (2000) 6105
\bibitem{r22b} P.M. Duxbury, D.J. Jacobs, M.F. Thorpe, C. Moukarzel,
Phys. Rev. E{\bf 59} (1999) 2084
\bibitem{Bool1} P. Boolchand and W.J. Bresser, Phil. Mag. B{\bf 80}
(2000) 1757
\bibitem{r261} L.F. Gladden and S.R. Elliott, Phys. Rev. Lett. {\bf 59}
(1987) 908
\bibitem{r262} A. Pradel, V. Michel-Lledos, M. Ribes and
E. Eckert, Chem. Mater. {\bf 5} (1993) 377
\bibitem{r263} E. Zintl, K. Loosen, Z. Phys. Chem. Leipzig {\bf 174}
(1935) 301
\bibitem{r264} M. Tenhover, M.A. Hazle, R.K. Grasseli,
Phys. Rev. B{\bf 29} (1984) 6732
\bibitem{r265} H. Eckert, J. Kennedy, A. Pradel and M. Ribes,
J. Non-Cryst. Solids {\bf 113} (1989) 187
\bibitem{r266} G. Dittmar, H. Schafer, Acta. Crysta. B{\bf 32} (1976) 2726
\bibitem{Tersoff} D.S. Franzblau and J. Tersoff, Phys. Rev. Lett. {\bf
68} (1992) 2172
\bibitem{Micoul} M. Micoulaut, J. Molec. Liquids {\bf 71} (1997) 107
\bibitem{KasapAsSe} T. Wagner, S.O. Kasap, M. Vlcek, A. Sklenar,
A. Stronski, J. Non-Cryst. Solids {\bf 227-230} (1998) 752
\bibitem{Raman} P. Boolchand, X. Feng, W.J. Bresser, J.Non-Cryst.
Solids {\bf 233} (2001) 348
\bibitem{FIC1} Z. Jinfeng and S. Mian-Zeng, {\em Materials for solid
state batteries}, ed. B.V.R. Chowdari and S. Radhakrishna (1986) p. 487
\bibitem{FIC2} S.W. Martin, Eur. J. Solid. St. Inorg. Chem. {\bf 28}
(1991) 163
\bibitem{condux1} S.W. Martin, J.A. Sills, J. Non-Cryst. Solids {\bf
135} (1991) 171
\bibitem{condux2} S. Sahami, S. Shea and J. Kennedy,
J. Electrochem. Soc. {\bf 132} (1985) 985
\bibitem{Science} M. Zhang and P. Boolchand, Science {\bf 266} (1994) 1355
\bibitem{SSC} R. Kerner and J.C. Phillips, Solid State Comm. {\bf 117}
(2000) 47
\bibitem{r341} O. Anderson, D. Stuart, J. Am. Ceram. Soc. {\bf 37}
(1954) 573
\bibitem{r342} D. Ravaine, J.L. Souquet, Phys. Chem. Glasses {\bf 18}
(1977) 27
\bibitem{Barrio} L.F. Perondi, R.J. Elliott, R.A. Barrio and K. Kaski,
Phys. Rev. B{\bf 50} (1994) 9868
\bibitem{Qn} J.F. Stebbins, J. Non-Cryst. Solids {\bf 106} (1988) 359
\bibitem{r343} see for example {\em Structure, Dynamics and Properties
of Silicate Melts}, Reviews in Mineralogy {\bf 32} (1995) 
\bibitem{r344} R. Narayanan, Phys. Rev. B{\bf 64} (2001) 134207
\bibitem{Bool2} P. Boolchand, M.F. Thorpe, Phys. Rev. B{\bf 50} (1994)
10366
\bibitem{CMV} F. Chaimbault, M. Micoulaut, Y. Vaills, unpublished
\bibitem{r346} J.F. Stebbins, P.W. MacMillan, J. Non-Cryst. Solids
{\bf 160} (1993) 116

\bibitem{ES1} A. Pradel, M. Ribes, Solid St. Ionics {\bf 18-19} (1986)
351
\bibitem{ES2} A. Pradel, M. Ribes, J. Non-Cryst. Solids {\bf 131-133}
(1991) 1063
\bibitem{Pradel} A. Pradel, J. Taillades, M. Ribes and H. Eckert,
J. Non-Cryst. Solids {\bf 188} (1995) 75
\bibitem{Vaills} Y. Vaills, G. Hauret, Y. Luspin, J. Non-Cryst. Solids
{\bf 286} (2001) 224
\end{thebibliography}
\end{document}